# Rapid and sensitive quantification of C3- and C6-phosphoesters in starch by fluorescence-assisted capillary electrophoresis

*Jeremy Verbeke[1], Christophe Penverne[1], Christophe D'Hulst[2], Christian Rolando[1], Nicolas Szydlowski[1,3]\**

[1] Univ. Lille, CNRS, USR 3290 – MSAP – Miniaturisation pour la Synthèse l'Analyse et la Protéomique, F-59000 Lille, France.

[2] Univ. Lille, CNRS, UMR8576 – UGSF – Unité de Glycobiologie Structurale et Fonctionnelle, F-59000 Lille, France.

[3] Present address: IFMAS – Institut Français des Matériaux Agrosourcés, 60 Avenue du Halley, F-59650 Villeneuve d'Ascq, France.

\* To whom correspondence should be addressed: Email: nicolas.szydlowski@univ-lille1.fr. Tel: +33-32-0434718

## ABSTRACT

Phosphate groups are naturally present in starch at C3- or C6-position of the glucose residues and impact the structure of starch granules. Their precise quantification is necessary for understanding starch physicochemical properties and metabolism. Nevertheless, reliable quantification of Glc-3-P remains laborious and time consuming. Here we describe a capillary electrophoresis method for simultaneous measurement of both Glc-6-P and Glc-3-P after acid hydrolysis of starch. The sensitivity threshold was estimated at the fg scale, which is compatible with the analysis of less than a μg of sample. The method was validated by analyzing antisense potato lines deficient in SBEs, GWD or GBSS. We show that Glc-3-P content is altered in the latter and that these variations do not correlate with modifications in Glc-6-P content. We anticipate the method reported here to be an efficient tool for high throughput study of starch phosphorylation at both C3- and C6-position.

1. INTRODUCTION

Starch accumulates as insoluble semi crystalline granules in plant cells as the result of the combination of two glucose polymers, amylopectin and amylose. In both polymers, glucose residues are linked together by α(1→4) and α(1→6) O-glycosidic bounds (Buléon, Colonna, Planchot & Ball, 1998). To date, the only known covalent modification of starch *in planta* is the phosphorylation of glucose residues (Blennow, Nielsen, Baunsgaard, Mikkelsen & Engelsen, 2002). Interestingly, most of the organic phosphate is found within the amylopectin fraction (Abe, Takeda & Hizukuri, 1982). At the glucose unit level, phosphorylation occurs at C6- or C3- position and traces of D-glucose-2-phosphate (Glc-2-P) were also reported in potato starch (Hizukuri, Tabata, Kagoshima & Nikuni, 1970; Tabata & Hizukuri, 1971). Phosphate content in starch varies remarkably according to the botanical origin. Among common starches, potato starch displays one of the highest phosphate contents (i.e. up to 0.09 %), second after shoti starch (*Curcuma zedoaria*) which contains 0.18 % of organic phosphate (BeMiller & Whistler, 2009).

In plants, two enzymes, the alpha-glucan, water dikinase, GWD and the phosphoglucan, water dikinase, PWD, phosphorylate starch at C6- or C3- position of the glucose residues, respectively (Ritte, Heydenreich, Mahlow, Haebel, Kötting & Steup, 2006). Inactivation of these enzymes in *Arabidopsis* leads to over accumulation of transitory starch, indicating impaired degradation at night (Kötting, Pusch, Tiessen, Geigenberger, Steup & Ritte, 2005; Yu et al., 2001). Phosphate groups were proposed to alter the crystallinity of amylopectin (Blennow & Engelsen, 2010). Such an amorphisation facilitates the hydrolases involved in starch breakdown to access glucose polymers at the granule surface (Blennow & Engelsen, 2010). In the *gwd* and *pwd* mutants, defective phosphorylation prevents these enzymes to bind their substrates and consequently reduces glucose remobilization (Kötting, Pusch, Tiessen, Geigenberger, Steup & Ritte, 2005; Yu et al., 2001). However, it was later proposed that GWD is essential for the synthesis of starch during the day (Hejazi, Mahlow & Fettke, 2014), while its function in degradation at night appears to be rather limited (Skeffington, Graf, Duxbury, Gruissem & Smith, 2014). In line with this, conformational study of C3- or C6- phosphorylated maltose indicate that 6-O-phosphate does not cause major steric changes and naturally fits within amylopectin double helices (Hansen et al., 2009). Conversely, 3-O-phosphate groups destabilize the conformational equilibrium of α(1→4) linkages, likely promoting amylopectin amorphisation (Blennow & Engelsen, 2010; Hansen et al., 2009).

While Glc-6-P can easily be determined by enzymatic assay, a number of methods were implemented for measuring Glc-3-P content in starch. These include colorimetric assay of total phosphate with subsequent subtraction of the Glc-6-P content, high performance anion exchange chromatography with pulsed-amperometric detection (HPAEC-PAD), hydrophilic interaction liquid chromatography-mass spectrometry (HILIC MS) and $^{31}$P NMR analysis (Blennow, Bay-Smidt, Olsen & Møller, 1998; Carpenter, Joyce, Butler, Genet & Timmerman-Vaughan, 2012; Morrison, 1964; Muhrbeck & Tellier, 1991). Most of these methods are time consuming and require hundreds of milligrams of starch. Nevertheless, HILIC MS has allowed for high throughput analysis of samples at the scale of the microgram (Carpenter, Joyce, Butler, Genet & Timmerman-Vaughan, 2012). Apart from $^{31}$P NMR which allows estimation of glucose phosphoesters in native starch, the other methods (i.e. colorimetric, HPAEC-PAD and mass spectrometry) require acid hydrolysis of the polymers (Carpenter, Joyce, Butler, Genet & Timmerman-Vaughan, 2012). However, Glc-3-P is dephosphorylated in acid conditions, resulting in the overestimation of the ratio Glc-6-P / Glc-3-P although the use of TFA (trifluoroacetic acid) was found to be more preservative than HCl treatment (Carpenter, Joyce, Butler, Genet & Timmerman-Vaughan, 2012; Haebel, Hejazi, Frohberg, Heydenreich & Ritte, 2008).

The present study reports on the use of capillary electrophoresis with laser induced fluorescence (LIF) detection for simultaneous measurement of Glucose, Glc-3-P and Glc-6-P after TFA hydrolysis of starch. We demonstrate through spiking experiments and alkaline phosphatase treatment that APTS (8-aminopyrene-1,3,6-trisulfonic acid)-derivatized glucose and glucose phosphoesters can be separated and determined specifically. We estimated the extent of dephosphorylation occurring concomitantly to starch hydrolysis for accurate measurement of Glc-3-P and applied the method to study C3- and C6- phosphorylation in transgenic potato lines deficient in starch branching enzymes (SBEs), the glucan water dikinase (GWD) or the granule bound starch synthase (GBSS) as well as in Arabidopsis, maize and parsnip. We show that Glc-3-P and Glc-6-P contents are affected in a non-linear relationship in the antisense potato lines although C6- and C3- phosphorylation are known to result from sequential enzymatic activities. The sensitivity of the method and the dynamic range of the LIF detector allowed for simultaneous integration of Glucose, Glc-3-P and Glc-6-P peaks in Arabidopsis and parsnip starches (containing approximately 0.5 phosphate ester / thousand glucose residues) in a single run. On the other hand, only Glc-6-P and glucose could be determined in maize starches containing ten times less phosphate.

## 2. MATERIALS AND METHODS

### 2.1. Starches and plant material

Starches from GBSS- (S11-51.1) (Kozlov, Blennow, Krivandin & Yuryev, 2007), GWD- (Dia H924-14.3, Dia H924-28.4) (Kozlov, Blennow, Krivandin & Yuryev, 2007; Viksø-Nielsen, Blennow, Jørgensen, Kristensen, Jensen & Møller, 2001) and BE- (Dia H944-3.3, Dia H944-6.1, Dia H944-14.5, Dia H944-15.1) (Blennow et al., 2005) suppressor lines were a gift from Pr. Andreas Blennow, Department of Plant and Environmental Sciences, University of Copenhagen, Denmark. Dried rhizomes of *Curcuma zedoaria* (shoti) were purchased in December 2011 on a local market near Kochi, Kerala, India. Maize and waxy maize starches were obtained from Roquette (Lestrem, France) and parsnip roots were purchased from a local grocer in Lille, France. Arabidopsis plants were cultivated in a greenhouse under a 16-h-light / 8-h-dark photoperiod at 23°C (day) / 20°C (night), 70 % humidity and a light intensity of 120 µE $m^{-2}$ $s^{-1}$.

### 2.2. Starch extraction

Arabidopsis starch was extracted from 21-day-old leaves according to (Delvalle et al., 2005). Shoti rhizomes were rehydrated at 4°C during 12 h before starch extraction. Parsnip roots and shoti rhizomes were ground with a blender in 200 mL of ultrapure water. The sample was filtered through a nylon net (100 µm mesh) and left for sedimentation of starch granules during 3 h. The supernatant was then removed and sedimented starch was resuspended in 500 mL of ultrapure water. Starch suspensions were subsequently washed 3 times with 1 L of ultrapure water and stored in 20 % ethanol at 4 °C or dried for long term storage.

### 2.3. Acid hydrolysis of starch polymers

Hydrolysis of starch was adapted from Carpenter, 2012 (Carpenter, Joyce, Butler, Genet & Timmerman-Vaughan, 2012). 1 mg whole starch was dispersed in 100 µL 2 M TFA prior to be incubated at 95 °C for the indicated period of time. Twenty μL aliquots of hydrolysates were then diluted with 200 μL of ultrapure water and subsequently dried at room temperature during 5 h in a speedvac. A second wash and 5 h drying step were performed after the addition of 200 μL ultrapure water. Aliquots, containing 200 μg dried sample, were then stored at -20°C prior to be used for APTS derivatization.

### 2.4. FACE analysis

Glc and Glc-6-P commercial standards were purchased from Sigma-Aldrich, France. Custom synthesis of Glc-3-P as well as its purity and identity control by $^1$H-NMR and LC-MS were purchased from ChiroBlock, Wolfen, Germany. TFA-treated samples and standards were labeled by adding 2 μL of 1 M sodium cyanoborohydride (Sigma-Aldrich, France) / THF (tetrahydrofurane) and 2 μL of 200 mM APTS (Sigma-Aldrich, France) / 15 % acetic acid and a subsequent overnight incubation at 42 °C. Electrophoresis was performed on reverse polarity with a Beckman Coulter PA800 plus instrument (Sciex Separations, Les Ulis, France) equipped with a laser induced fluorescence detector. Separation was performed in a 50 μm I.D., 375 μm O.D. bare fused silica capillary of 60.2 cm in length (Sciex Separations, Les Ulis, France). Sample injection was carried out at 0.5 psi during 10 sec and separation was achieved at 30 kV in carbohydrate separation gel buffer (Sciex Separations, Les Ulis, France), pure or diluted 1/3 in ultrapure water.

### 2.5. Alkaline phosphatase treatment

Shoti starch hydrolysates were submitted to APTS derivatization as described above prior to addition of 100 μL of 0.1 M sodium carbonate, pH 9.9 containing 1,2U/ml Bovine alkaline Phosphatase (Sigma-Aldrich, France). Samples were incubated for 1 hour at 37°C and analyzed by capillary electrophoresis as described above.

### 2.6. Spiking experiments

Glc-3-P and Glc-6-P peak areas in shoti starch were determined with the use of the software 32Karat and quantification of both compounds was carried out using known amounts of Glc-3-P or Glc-6-P. The shoti starch hydrolysate was diluted 2 times and analyzed by FACE after addition of half of the Glc-3-P or Glc-6-P amounts measured in the sample.

## 3. RESULTS AND DISCUSSION

### 3.1. Description of the method

**Scheme 1.** Principle of the method.

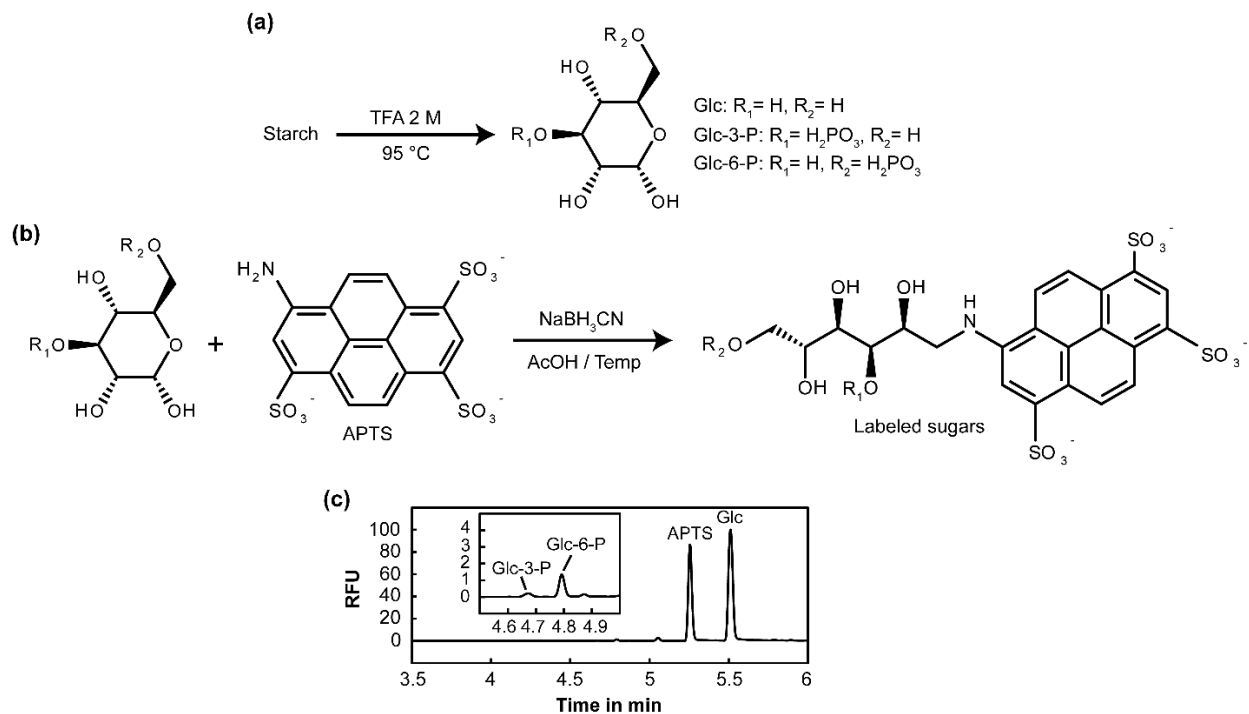

(a) Scheme of the acid hydrolysis of starch with the use of trifluoroacetic acid (TFA) and heating. A mix of glucose (Glc), D-glucose-3-phosphate (Glc-3-P), D-glucose-6-phosphate (Glc-6-P) is produced from the complete digestion of starch.

(b) Labeling reaction of the hydrolysis products by reductive amination with 8-aminopyrene-1,3,6-trisulfonic acid (APTS).

(c) Electrophoregram of the labeled products obtained by FACE. The inset displays the signal enlarged between 4.5 and 5 min. Compounds are detected between 4.5 and 5.5 min in the order: Glc-3-P, Glc-6-P, free APTS, Glc.

Acid hydrolysis of starch was coupled to reductive amination of released glucose and glucose phosphoesters with 8-aminopyrene-1,3,6-trisulfonic acid (APTS) prior to analysis by fluorescence assisted capillary electrophoresis (FACE) (Scheme 1). The use of charged fluorophores has for long proven successful in the determination of mono- and poly-saccharides (O'Shea, Samuel,

Konik & Morell, 1998). In particular, their separation in gel electrophoresis is facilitated and the fluorescence detection offers high sensitivity. The detection thresholds of APTS-Glc-3-P and APTS-Glc-6-P conjugates were determined by analyzing serial dilutions of the labeled standards (Supplemental figure S1). Resolved peaks with good signal to noise ratios were observed down to a sample concentration of 2 pg / µL and a 10 sec injection at 0.5 psi (Supplemental figure S1). The corresponding injection volume was 10 nL as determined with the Poiseuille equation: $V = (\Delta P d^4 \pi t) / (128 \eta L)$, where V is the injected volume in nL, $\Delta P$ is the pressure drop within the capillary in Pa, $d$ is the internal diameter of the capillary in m, $t$ is the time of applied pressure in sec, $\eta$ is the fluid viscosity in Pa.sec$^{-1}$ and $L$ is the length of the capillary in m. Thus, the estimated threshold of sensitivity was at the fg level, compatible with phosphoester measurement in less than one µg of starch. Electrophoretic separation of the hydrolysis products from potato starch, as well as free APTS (i.e. APTS molecules that did not react with sugars), was achieved in the order: Glc-3-P, Glc-6-P, APTS, Glc (Scheme 1c, Figure 1a). Both Glc-3-P and Glc-6-P peaks disappeared after treatment with alkaline phosphatase (Figure 2a). Moreover, spike assessment of both compounds was performed by adding a known amount of the respective standards to starch hydrolysates (Figure 2b). Increases of the corresponding peak areas were proportional to the amount of exogenous standards, thus corroborating the peak identities and the absence of co-eluted compounds. The method was further validated by analyzing already characterized antisense potato starches with altered phosphorus contents (Blennow et al., 2005; Kozlov, Blennow, Krivandin & Yuryev, 2007; Viksø-Nielsen, Blennow, Jørgensen, Kristensen, Jensen & Møller, 2001) as well as the highly phosphorylated shoti starch (Ithaca & Maywald, 1968)(Figure 1b). Consistent with previous descriptions, Glc-3-P and Glc-6-P peak areas were increased in both shoti starch and the antisense potato line asBE, whereas they were significantly decreased in starch from the asGWD line (Figure 1b).

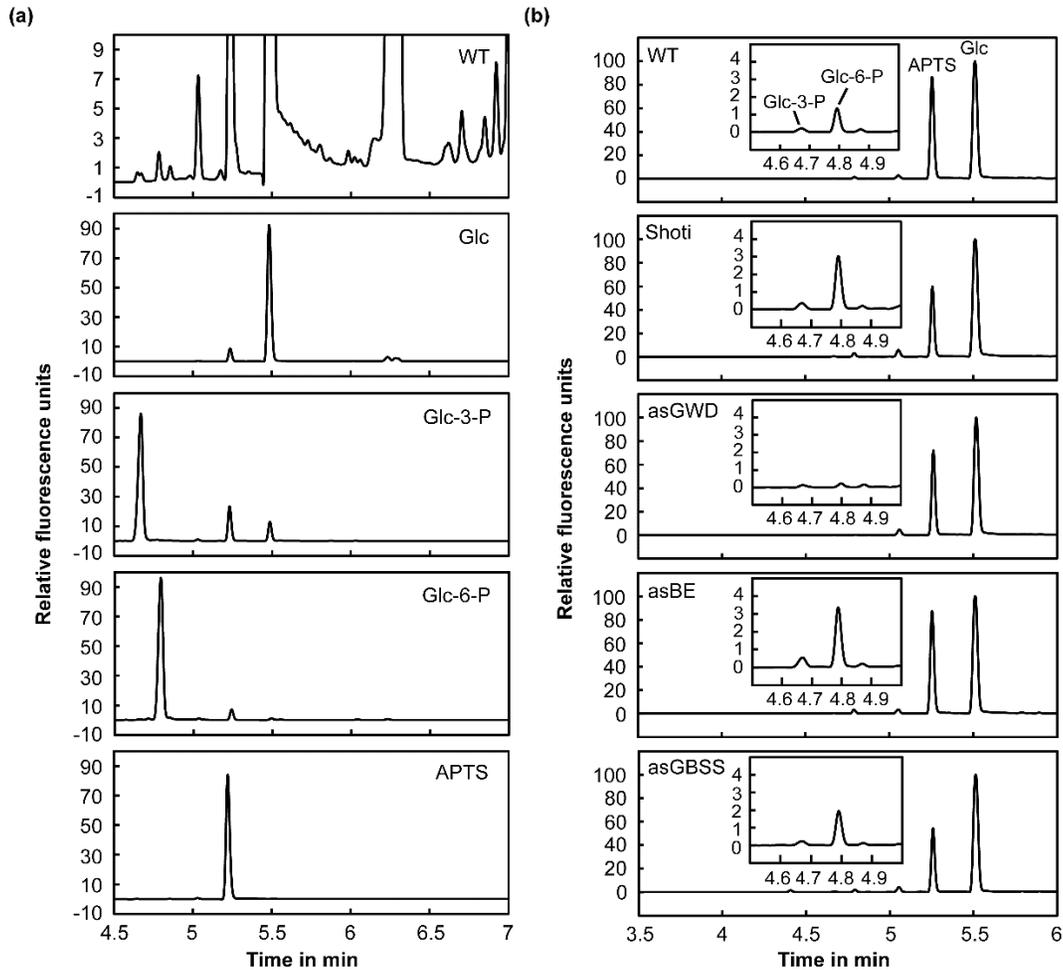

**Figure 1.** FACE analysis of TFA hydrolyzed starches and the corresponding standards.

(a). Electrophoretic profiles of hydrolyzed potato (cv. Dianella) starch (WT) as well as the standard compounds, glucose (Glc), D-glucose-3-phosphate (Glc-3-P), D-glucose-6-phosphate (Glc-6-P) and 8-aminopyrene-1,3,6-trisulfonic acid (APTS). Samples were submitted to acid hydrolysis during 60 min in TFA 2 M prior to APTS derivatization. Fifty µL samples containing 1 µg of derivatized products were used for hydrodynamic injection at 0.5 psi during 3 sec, corresponding to an injection volume of 3 nL.

(b). Hydrolysis products of wild type (WT) potato (cv. Dianella) or "shoti" starches and from the antisense potato lines deficient in glucan water dikinase (asGWD, Dia H924-14.3), starch branching enzyme (asBE, Dia H944-3.3) or granule bound starch synthase (asGBSS, S11-51.1). Starches were treated and analyzed as in (a). Insets display enlarged profiles between 4.5 and 5 min. The peaks of Glc-3-P, Glc-6-P, free APTS and glucose (Glc) are indicated.

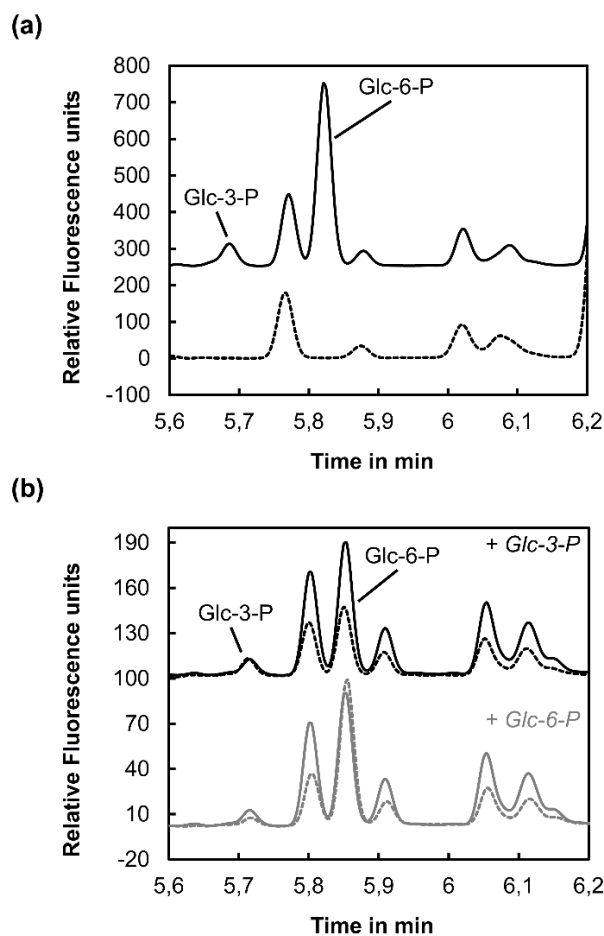

**Figure 2:** Glc-3-P and Glc-6-P peak identity.

(a). Electrophoretic traces of APTS-derivatized hydrolysis products from shoti starch, treated (dotted line) or not (continuous line) with alkaline phosphatase. The profile of the treated sample was plotted with a y-offset of 250 RFU (Relative Fluorescence Units). The peaks of Glc-3-P and Glc-6-P are indicated.

(b). Hydrolysis products from shoti starch (continuous lines) were spiked (dotted lines) with the standards Glc-3-P (black lines) or Glc-6-P (grey lines). Glc-3-P and Glc-6-P were quantified in shoti starch hydrolysates using known amounts of standards. Samples were diluted twice prior to addition of half the Glc-3-P or Glc-6-P contents measured in the samples (i.e. 0.68 pg and 9.39 pg, respectively). Top electrophoregrams were plotted with a y-offset of 100 RFU. The peaks of Glc-3-P and Glc-6-P are indicated.

The TFA-treated standards Glc-3-P and Glc-6-P displayed an additional peak of glucose (Figure 1a). However, this peak was not observed in the commercial batches of standards when the TFA treatment is omitted (Supplemental figure S2), indicating that dephosphorylation takes place concomitantly to starch hydrolysis. In order to assess the effect of TFA on the phosphoester bond integrity, time course experiments were performed at 95 °C (Figures 3 and 4). Dephosphorylation of the standards started rapidly as illustrated by glucose releases of 4 % and 1.6 % after 20 min from Glc-3-P and Glc-6-P, respectively (Figure 3). However, Glc-6-P was the most stable compound since 96.5 % of the standard remained after 180 min of TFA treatment (Figure 3b and 3c). Conversely, dephosphorylation of Glc-3-P increased linearly over 140 min prior to reach a plateau at approximately 40 % (Figure 3a and 3c). The dynamic range of the LIF detector was tuned for simultaneous integration of Glc, Glc-3-P and Glc-6-P peaks in starch hydrolysates and similar time course was performed with wild type potato (*Solanum tuberosum* cv. Dianella) starch (Figure 4). Peak areas of glucose, Glc-3-P and Glc-6-P were monitored during 180 min (Figure 4a, 4b and 4c). While the amount of released glucose reached a plateau after 100 min, Glc-3-P and Glc-6-P peak areas increased over the complete time course (Figure 4a, 4b and 4c). The ratio between C3- and C6-phosphoesters also increased rapidly and peaked after 40 min prior to decrease over 100 min (Figure 4d). Correction factors were estimated from the time course dephosphorylation of the standard compounds (Figure 3) and applied to the amounts of phosphoesters released from starch (Figure 4d). After correction for dephosphorylation, the proportion of Glc-3-P was stable after 80 min (Figure 4d). Thus, in the rest of the study, we performed phosphoester measurements after 80 min of TFA hydrolysis and with the use of the correction factors 1.24 and 1.02 to adjust the Glc-3-P and Glc-6-P peak areas, respectively.

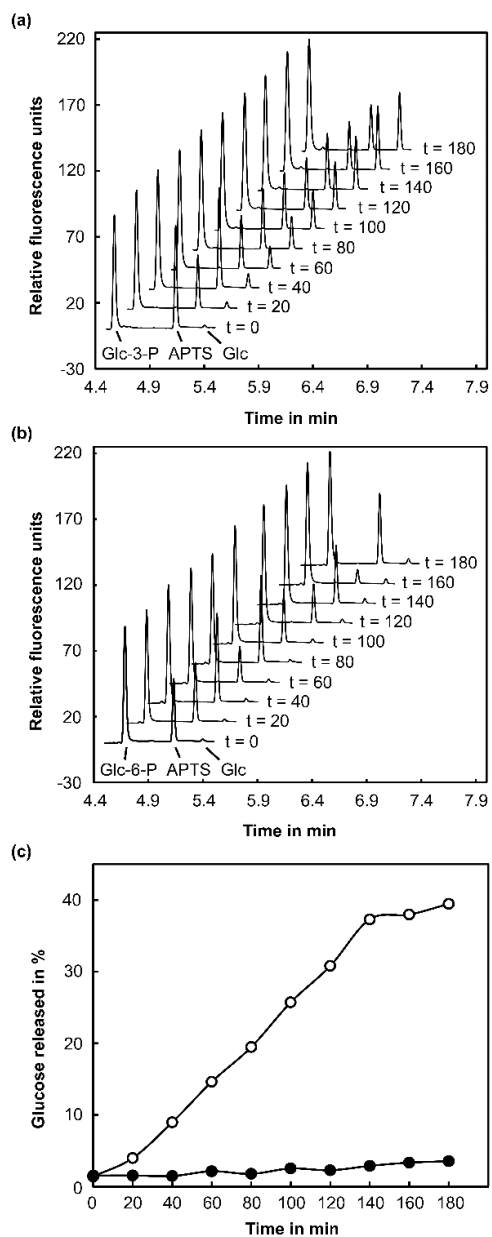

**Figure 3.** Time course dephosphorylation of the standards Glc-3-P and Glc-6-P submitted to TFA treatment.

(a) Electrophoretic traces of Glc-3-P submitted to TFA treatment over a 180 min time course. Aliquots were taken at the indicated time points prior to FACE analysis. Electrophoregrams were plotted with incremental x- and y-offsets of 0.3 min and 15 RFU (Relative Fluorescence Units), respectively. The peaks corresponding to Glc-3-P, APTS and Glc are indicated.

(b) As in (a) with the use of the standard compound Glc-6-P.

(c) Glucose released from Glc-3-P (open circles) or Glc-6-P (closed circles) during TFA treatment. Released glucose is expressed in percentage compared with the amount of remaining intact standard.

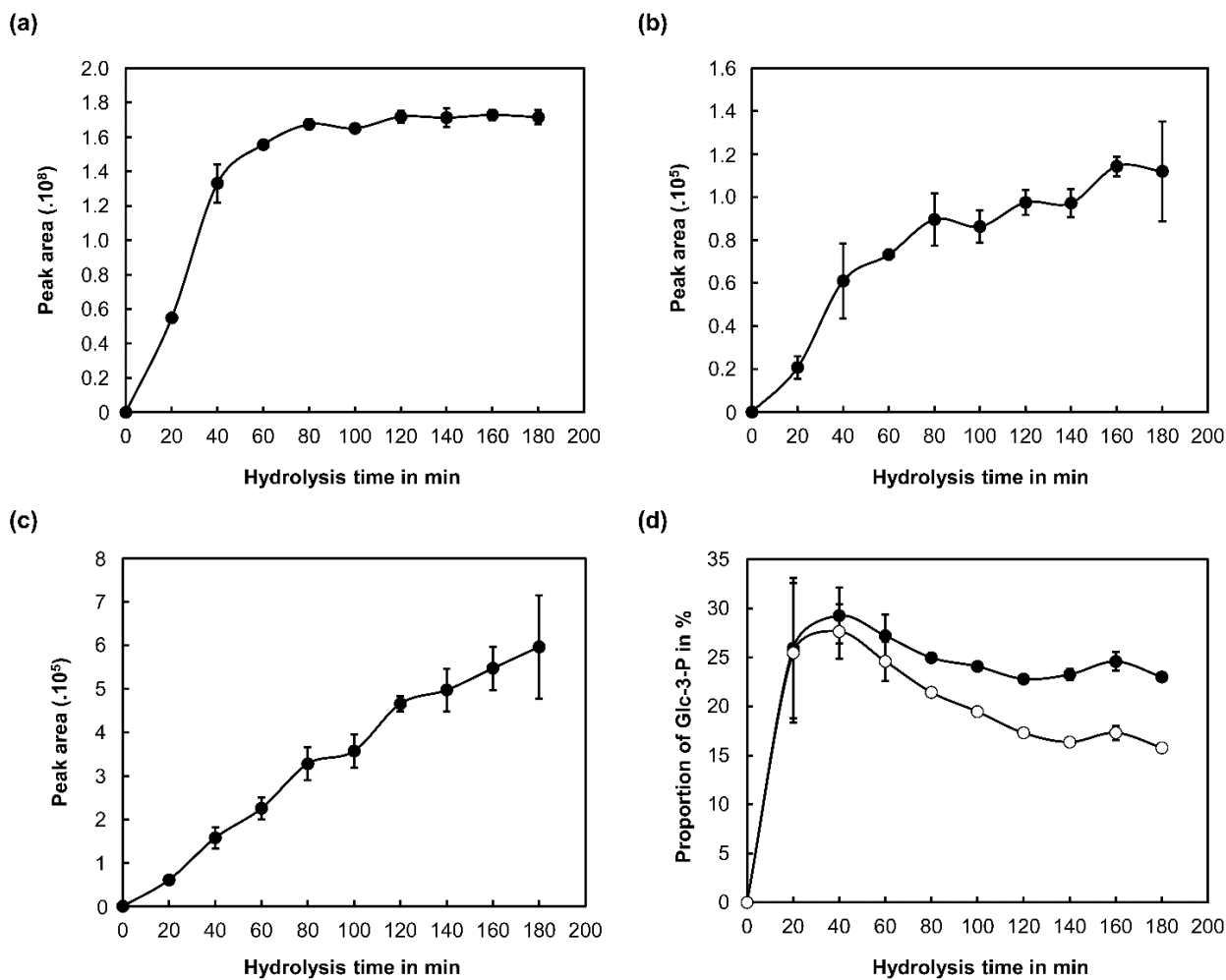

**Figure 4.** Time course hydrolysis of potato starch.

Amounts of Glc (a), Glc-3-P (b) and Glc-6-P (c) released from WT potato (*cv.* Dianella) starch submitted to TFA hydrolysis were monitored during 180 min. Starch was submitted to TFA treatment and aliquots were taken at the indicated time points. Samples were analyzed by FACE after derivatization with APTS and the corresponding peak areas were plotted versus time. Errors bars represent the standard deviation of two independent time course hydrolysis. (d) Proportion of Glc-3-P expressed in percent of total phosphate. Open and close circles correspond to observed and corrected values, respectively. Correction factors were obtained from the time course dephosphorylation of the standards Glc-3-P and Glc-6-P displayed in Figure 3c.

### 3.2. Phosphate determination in starches from GWD-, BE- and GBSS-antisense potato lines

C3- and C6-phosphoesters were measured by FACE in potato (cv. Dianella) starch and in the antisense lines asGWD (Dia H924-14.3, Dia H924-28.4), asBE (Dia H944-3.3, Dia H944-6.1, Dia

H944-14.5, Dia H944-15.1), asGBSS (S11-51.1), as well as PAPS (Potato amylopectin starch) which were already described elsewhere (Blennow et al., 2005; Kozlov, Blennow, Krivandin & Yuryev, 2007; Viksø-Nielsen, Blennow, Jørgensen, Kristensen, Jensen & Møller, 2001) (Table 1). In order to test the accuracy of the FACE method, the latter was compared with $^{31}$P-NMR spectroscopy (Supplemental Figure S3) (Kasemsuwan & Jane, 1996). This method does not require acid hydrolysis of starch thus preventing dephosphorylation during sample preparation. The absolute phosphoester contents determined by NMR in Dianella starch were comparable to those measured by FACE (Table 1 and Supplemental Figure S3). Moreover, the proportion of C3 phosphoesters assayed by NMR and FACE were 17.92 % and 19.13 %, respectively. These data corroborated the use of the correction factors 1.24 and 1.02 after 80 min of TFA hydrolysis for accurate determination of the ratio Glc-6-P / Glc-3-P. Several studies reported that C3 phosphoesters contribute to approximately 30 % of the total phosphate content in potato starch (Blennow, 2015; Hizukuri, Tabata, Kagoshima & Nikuni, 1970; Tabata & Hizukuri, 1971). On the other hand, starch phosphorylation is influenced by plant growth conditions. In particular, the use of fertilizer significantly impacts starch phosphorylation (Jacobsen, Madsen, Christiansen & Nielsen, 1998) and may explain this discrepancy. Moreover, long term storage of starch samples is likely leading to dephosphorylation and thus to the underestimation of the phosphate contents. This is illustrated by the overall lower phosphate contents measured in this study when compared with previous descriptions of the same samples (Table 1) (Blennow et al., 2005; Kozlov, Blennow, Krivandin & Yuryev, 2007).

Antisense inhibition of the glucan water dikinase (GWD) leads to higher amylose content and a decrease in Glc-6P content (Viksø-Nielsen, Blennow, Jørgensen, Kristensen, Jensen & Møller, 2001). Our results were consistent with these previous descriptions since total phosphate was significantly lower in starches from Dia H924-14.3 and Dia H924-28.4 compared to the wild type (Table 1). The content of Glc-6-P was decreased by 84.62 % and 83.71 % in lines Dia H924-14.3 and Dia H924-28.4, respectively (Table1). In addition, our study revealed that C3 phosphorylation is also drastically affected in these lines which only accumulate 26.53 % of the wild type content (Table 1). Conversely, the proportion of Glc-3-P (expressed in percent of total phosphate) was significantly increased (Table 1).

**Table 1:** Phosphoester contents and proportion of Glc-3-P in starches from wild type and antisense potato lines (cv Dianella) as well as PAPS. Phosphoester contents were determined by FACE after 80 min of TFA hydrolysis and expressed per thousand of Glucose residues. The proportion of C3 phosphoesters was expressed in percent of total phosphate. The correction factors, 1.24 and 1.02, were applied to Glc-3-P and Glc-6-P, respectively, to correct for dephosphorylation concomitant to starch hydrolysis. The standard deviation of three experimental replicates is indicated. Significant differences compared to the wild type reference (cv Dianella) were determined with a student's $t$-test with $P \leq 0.001$ ([a]), $P \leq 0.02$ ([b]) or $P \leq 0.05$ ([c]). Previously described Glc-6-P contents ([d]) (Kozlov, Blennow, Krivandin & Yuryev, 2007) and ([e]) (Blennow et al., 2005) were converted from nmol / mg starch to relative proportions (‰) and provided for comparison.

| Genotype | Line | Total phosphate content (‰) | Content of Glc-3-P (‰) | Content of Glc-6-P FACE (‰) | Content of Glc-6-P Enzyme-linked assay (‰) | Proportion of C3-phosphoesters (%) |
|---|---|---|---|---|---|---|
| Wild type | *Solanum tuberosum* cv Dianella | 2.73 ± 0.21 | 0.52 ± 0.03 | 2.21 ± 0.19 | 3.67[d] | 19.13 ± 1.3 |
| asGWD | Dia H924-14.3 | 0.48 ± 0.04 [a] | 0.14 ± 0.03 [a] | 0.34 ± 0.03 [a] | 0.60[d] | 28.92 ± 1.5 [b] |
| asGWD | Dia H924-28.4 | 0.50 ± 0.03 [a] | 0.13 ± 0.03 [a] | 0.36 ± 0.01 [a] | ND | 27.53 ± 3.2 [b] |
| asGBSS | S11-51.1 | 3.66 ± 0.11 [a] | 0.49 ± 0.09 | 3.17 ± 0.02 [a] | 5.54[d] | 13.41 ± 2.1 [b] |
| Waxy | PAPS | 3.68 ± 0.12 [a] | 0.56 ± 0.07 | 3.12 ± 0.12 [a] | ND | 15.39 ± 1.9 [c] |
| Control | Dia H944 | ND | ND | ND | 5.07[e] | ND |
| asBE | Dia H944-3.3 | 6.49 ± 0.51 [a] | 1.29 ± 0.18 [a] | 5.19 ± 0.34 [a] | 13.90[e] | 20.00 ± 1.2 |
| asBE | Dia H944-6.1 | 6.03 ± 0.19 [a] | 1.20 ± 0.13 [a] | 4.83 ± 0.07 [a] | 13.76[e] | 19.93 ± 1.4 |
| asBE | Dia H944-14.5 | 7.86 ± 1.23 [a] | 1.35 ± 0.22 [a] | 6.51 ± 1.17 [a] | 16.52[e] | 17.41 ± 3.0 |
| asBE | Dia H944-15.1 | 7.10 ± 0.52 [a] | 1.39 ± 0.08 [a] | 5.72 ± 0.53 [a] | 14.59[e] | 19.74 ± 1.8 |

Based on *in vitro* studies of the corresponding recombinant proteins, phosphorylation at C3 position by the phosphoglucan water dikinase (PWD) is thought to be dependent on the anterior activity of GWD, which phosphorylates at C6 position (Hejazi et al., 2008). Thus, one possible explanation is that alteration of GWD does not impact the subsequent activity of PWD in a linear relationship. Moreover, the modification of starch polymer composition (i.e. increase in amylose content) in these lines may also differently influence phosphorylation at C3 positions by PWD.

In lines Dia H944-3.3, Dia H944-6.1, Dia H944-14.5 and Dia H944-15.1, expression of both isoforms of starch branching enzyme (i.e. SBE1 and SBE2) is altered (Blennow et al., 2005). This results in increased amylose and Glc-6-P contents as well as longer amylopectin chains (Blennow et al., 2005). Our data show that in addition to the Glc-6-P content, Glc-3-P is also increased in the four lines (Table 1). Increases in Glc-3-P and Glc-6-P are positively correlated and range from 131 to 167 % and 119 to 160 %, respectively (Table 1). Consequently, the proportion of C3-phosphoesters remains unchanged in the asBE lines compared to the wild type.

On the other hand, phosphate content was slightly increased in asGBSS (S11-51.1) and PAPS (Table 1). Intriguingly, this difference is due to the significant increase of Glc-6-P while Glc-3-P content remains unchanged leading to a slight diminution of the proportion of Glc-3-P (Table 1). Starch phosphoesters have been described as essentially located on amylopectin molecules (Abe, Takeda & Hizukuri, 1982). Furthermore, the biosynthesis of amylose is drastically reduced (i.e. decreased by about 98 % compared to the wild type) (Kozlov, Blennow, Krivandin & Yuryev, 2007) or abolished in asGBSS and PAPS, respectively. On the contrary, starch from the wild type contains approximately 31 % of amylose (Kozlov, Blennow, Krivandin & Yuryev, 2007). Therefore, normalization of the data to the total amount of glucose units may impact the variation of phosphoester contents in these samples. Assuming that phosphorus is essentially located on amylopectin molecules, we performed normalization of the phosphoester contents to the amylopectin concentration in wild type and asGBSS starches (Figure 5). Glc-6-P content was similar in both samples while Glc-3-P was significantly reduced by 33 % indicating that amylose biosynthesis and the establishment of proper starch phosphorylation at C3 position may be interconnected. A study from Buléon and collaborators (Buleon, Cotte, Putaux, d'Hulst & Susini, 2014) showed that phosphorus repartition is altered in starch granules of *amf* (*amylose free*) mutant potato starch, as seen by synchrotron X-ray microfluorescence mapping, corroborating that potential interconnection (Buleon, Cotte, Putaux, d'Hulst & Susini, 2014). Furthermore, a negative covariation between branching level and phosphate content was observed in starches of the asBE potato lines (Blennow et al., 2005).

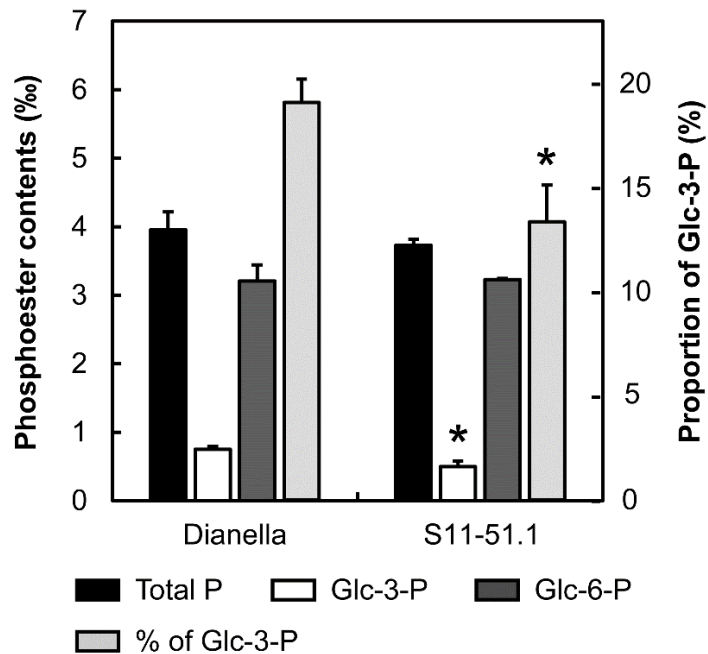

**Figure 5:** Phosphoester contents normalized to amylopectin concentration in potato (cv. Dianella) starch and the antisense line asGBSS (S11-51.1).

Glc-3-P (white bars), Glc-6-P (dark grey bars) and total phosphate (black bars) contents and the proportion of Glc-3-P (light grey bars) were determined by FACE after APTS-derivatization of starch hydrolysis products (Table 1). Values were normalized to amylopectin concentration (i.e. 68.9 % and 98.1 % in wild type and asGBSS, respectively (Kozlov, Blennow, Krivandin & Yuryev, 2007)). Glc-3-P, Glc-6-P and total phosphate contents were expressed ‰ glucose residues (primary y-axis) and the proportion of Glc-3-P was expressed in % of total phosphate (secondary y-axis). Errors bars represent the standard deviation of three experimental replicates. The asterisks indicate significant differences compared to the wild type reference according to a student's *t*-test with $P \leq 0.01$.

One proposed explanation is that GWD preferentially phosphorylates the longer unit chains of potato starch (Blennow et al., 2005; Mikkelsen, Baunsgaard & Blennow, 2004). Taken together, our data show that phosphorylation at C3-position is also affected in potato starch following inhibition of BE, GWD and GBSS and that these alterations are not correlated to modifications in Glc-6-P content. Although the biochemical basis of this phenomenon remains to be investigated, our study suggests that sequential activities of GWD and PWD does not follow a linear relationship and may be modulated by other factors such as starch branching level or amylose synthesis.

### 3.3. Phosphoester determination in low phosphate-starches

Potato and shoti are known to accumulate the most highly phosphorylated starches. We were prompted to apply the FACE method to starches with lower phosphate contents in order to evaluate its level of sensitivity. Similar to other cereals, maize starch contains traces of phosphoesters while that of Arabidopsis displays intermediate phosphate levels. Thus, capillary electrophoresis was performed on hydrolysis products from commercial maize- and Waxy maize-starches as well as *Arabidopsis thaliana* and parsnip (Supplemental Figure S4 and Table 2). The dynamic range of the LIF detector allowed for simultaneous integration of the glucose, Glc-3-P and Glc-6-P peaks in parsnip and Arabidopsis (Supplemental Figure S4). On the other hand, only Glc-6-P could be detected in maize starches at this sample concentration (Supplemental Figure S4). Increasing the latter led to saturation of the glucose peak, incompatible with phosphoester determination in a single run. The phosphate contents of maize and Arabidopsis starches (i.e. 0.03 ‰ and 0.6 ‰, respectively) and the low proportion of C3-phosphoesters in Arabidopsis (i.e. 12.3 %) were in agreement with previous reports (Kasemsuwan & Jane, 1996; Santelia et al., 2011).

**Table 2:** Phosphoester contents and proportion of Glc-3-P in starches from maize, Arabidopsis and parsnip. Phosphoester contents were determined by FACE and expressed per thousand of Glucose residues. The proportion of Glc-3-P was estimated after correction for dephosphorylation with the use of the factors 1.24 and 1.02 for Glc-3-P and Glc-6-P, respectively. The standard deviation of three (parsnip and Arabidopsis) or six (maize and waxy maize) experimental replicates is indicated.

| Common name | Species | Total phosphate content (‰) | Content of Glc-3-P (‰) | Content of Glc-6-P (‰) | Proportion of C3-phosphoesters (%) |
|---|---|---|---|---|---|
| Parsnip | *Pastinaca sativa* | 0.58 ± 0.03 | 0.070 ± 0.01 | 0.51 ± 0.04 | 12.27 ± 2.9 |
| Arabidopsis | *Arabidopsis thaliana* | 0.62 ± 0.04 | 0.072 ± 0.01 | 0.55 ± 0.03 | 11.50 ± 1.5 |
| Maize | *Zea mays* | 0.034 ± 0.013 | ND | 0.034 ± 0.013 | ND |
| Maize (Waxy) | *Zea mays (Waxy)* | 0.032 ± 0.012 | ND | 0.032 ± 0.012 | ND |

## 4. CONCLUSIONS

This study is the first report on the use of capillary electrophoresis for phosphate determination in starch. The method reported here is straightforward and offers major advantages including cost-effectivity, high sensitivity and is suitable for high throughput determination of both C3- and C6-phosphoesters. Thus, FACE-measurement of Glc-3-P and Glc-6-P can be coupled to other separation techniques such as starch polymer fractionation, either in their native form or after enzymatic hydrolysis. This will allow deciphering the repartition of both types of phosphoesters on starch glucose unit chains in relation with branching level. Moreover, the level of sensitivity reported here (i.e. at the scale of the fg) is promising for the study of starch phosphorylation even in organisms accumulating tiny amounts of starch.

Here we show that substantial level of dephosphorylation at C3 position occurs during starch hydrolysis with TFA at high temperature, although to a lesser extent than in the presence of HCl as this was previously described (Carpenter, Joyce, Butler, Genet & Timmerman-Vaughan, 2012; Haebel, Hejazi, Frohberg, Heydenreich & Ritte, 2008). We determined that TFA hydrolysis of starch during 80 min in combination with correction factors of 1.24 and 1.02 for Glc-3-P and Glc-6-P, respectively, is optimal for accurate determination of both phosphoesters. Thus, the ratio between C3- and C6-phosphorylation can be followed accurately allowing for dissection of the biochemical and metabolic interaction between both types of phosphorylation. Our study of antisense potato starches deficient in SBEs, GWD and GBSS showed that subtle phenotypes can be assessed, especially regarding the Glc-3-P content of starch. We show that variations in phosphate contents in these lines is not always due to correlated changes in Glc-3-P and Glc-6-P contents and that modulation of this ratio may also be related to starch structure and amylose biosynthesis.

**SUPPORTING INFORMATION**

Here we include the electrophoregrams of serial dilutions of the standard compounds Glc-3-P and Glc-6-P (Supplemental figure S1) as well as those showing the absence of glucose in the commercial batches of standards (Supplemental figure S2) that we did not included in the body of the manuscript. We also provide a $^{31}$P-NMR spectrum of potato α-limit dextrins and the FACE analysis of Arabidopsis, parsnip and maize starches. All of these data were obtained as described in the Experimental section of this manuscript.

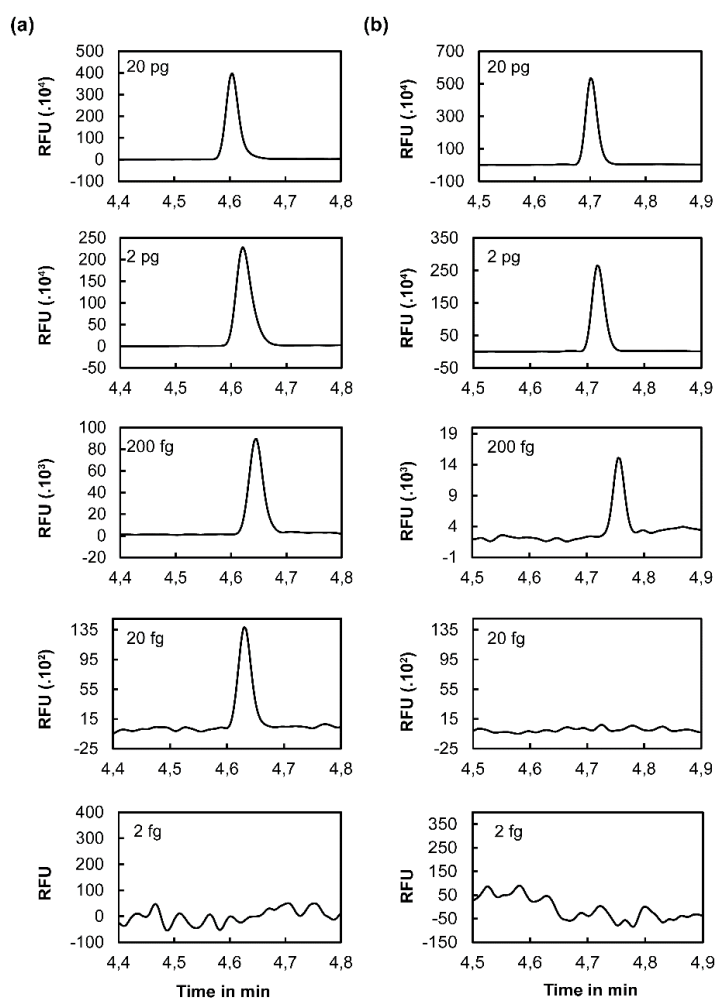

**Supplemental figure S1:** FACE analysis of the indicated amounts of APTS-Glc-3-P (a) and APTS-Glc-6-P (b).

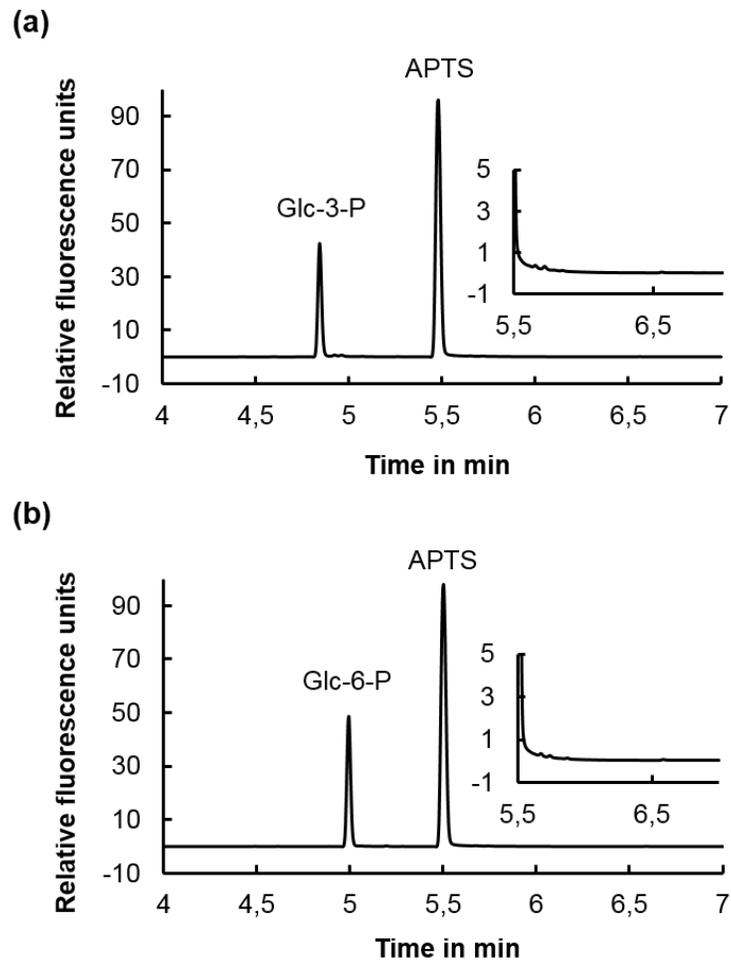

**Supplemental figure S2:** FACE analysis of the standard compounds Glc-3-P (a) and Glc-6-P (b) prior to TFA treatment. The insets show enlarged profiles at the time when glucose peak is detected, if present.

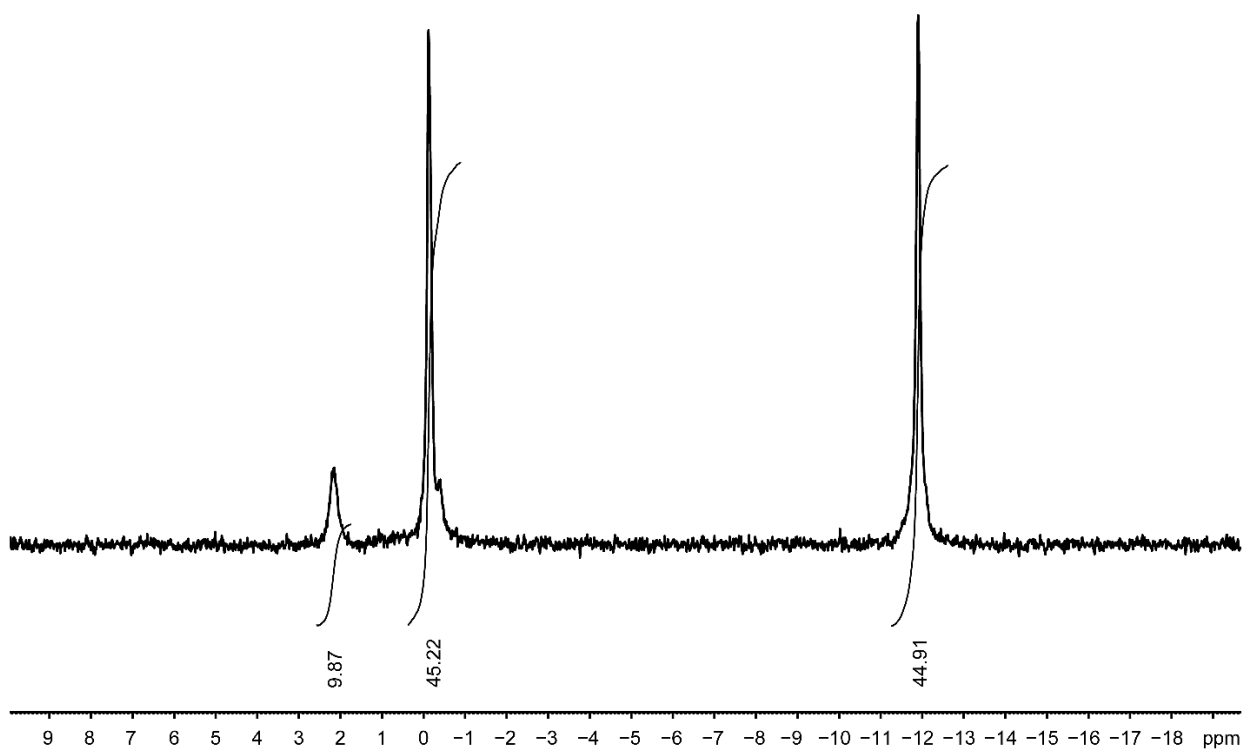

**Supplemental figure S3:** $^{31}$P-NMR spectrum of WT potato (*cv.* Dianella) α-limit dextrins. Two g of starch were submitted to α-amylolysis prior to $^{31}$P-NMR spectroscopy according to (Kasemsuwan & Jane, 1996) with the addition of NAD as an internal standard (10.5 mg). The spectrum was acquired with a Bruker Avance 300 spectrometer (Bruker Corporation, Billerica, MA) at a frequency of 125 MHz, flip angle 90° (10 μsec), sweep width 400 ppm, 65 k data points and a temperature of 298 °K. Glc-3-P (signal at 2.1 ppm) and Glc-6-P (signal at - 0.2 ppm) contents were calculated from the ratio of their respective peak areas and that of NAD (-12 ppm). Total phosphate content corresponds to the sum of Glc-3-P and Glc-6-P contents. The proportion of both phosphoesters was estimated from the ratio of their respective peak areas.

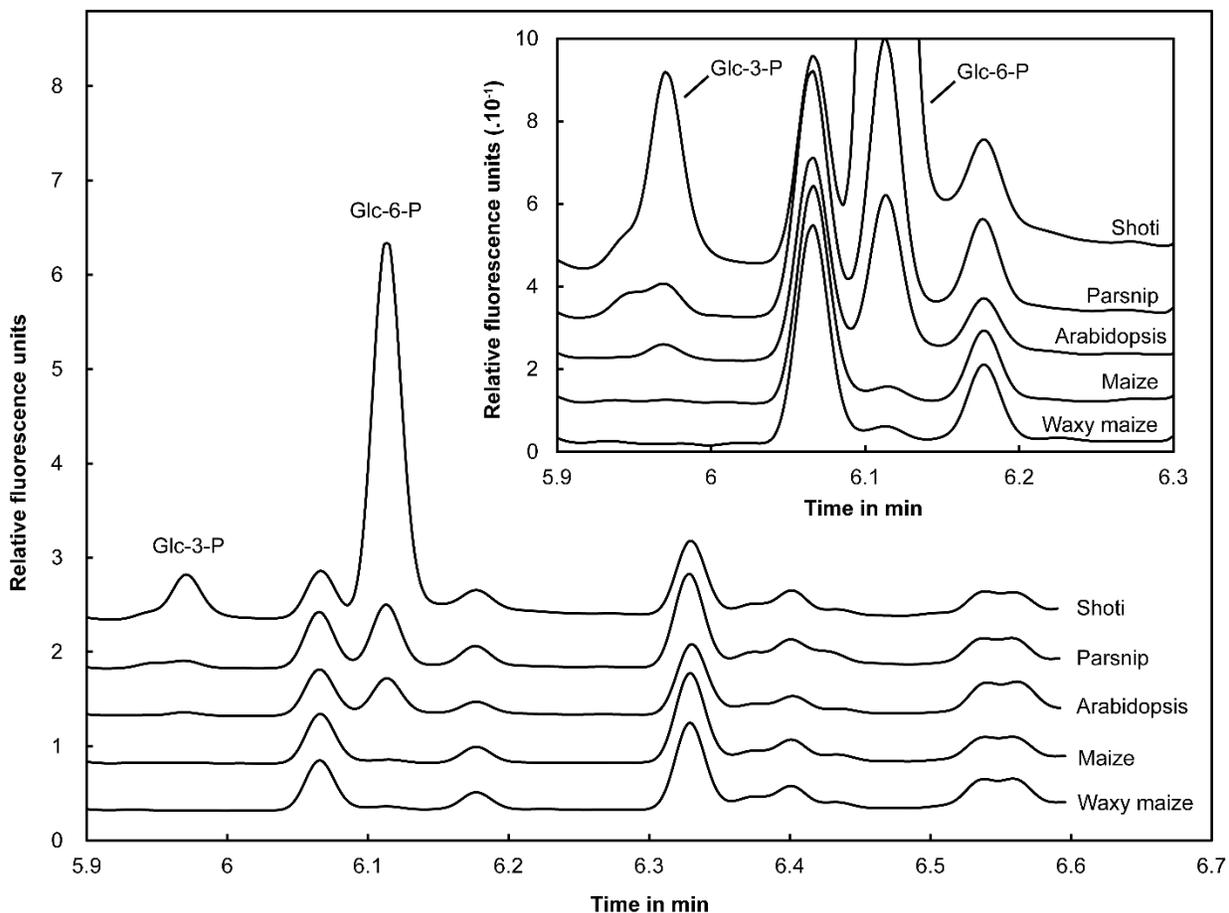

**Supplemental figure S4:** FACE analysis of starches containing trace- and intermediate-phosphate contents. One ng of APTS-derivatized hydrolysis products from shoti, parsnip, Arabidopsis, maize and waxy maize starches were analyzed. The inset displays enlarged profiles between 5.9 and 6.3 min. The peaks of Glc-3-P and Glc-6-P are indicated.

**ACKNOWLEDGMENTS**

This research was supported by ANR grants through the PDOC-2013 and the "investment for the future" programs (ANR-13-PDOC-0030-01 and ANR-10-IEED-0004-01, respectively). The authors thank Andreas Blennow for providing the wild type and antisense potato starches analyzed in this study. The authors also acknowledge the NMR platform of the University of Lille 1 for technical assistance with NMR-spectroscopy.

# REFERENCES

Abe, J.-I., Takeda, Y., & Hizukuri, S. (1982). Action of glucoamylase from aspergillus niger on phosphorylated substrate. *Biochimica et Biophysica Acta (BBA) - Protein Structure and Molecular Enzymology, 703*(1), 26-33.

BeMiller, J. N., & Whistler, R. L. (2009). *Starch: Chemistry and Technology*. Elsevier Science.

Blennow, A. (2015). Phosphorylation of the Starch Granule. In Y. Nakamura (Ed.). *Starch: Metabolism and Structure* (pp. 399-424). Tokyo: Springer Japan.

Blennow, A., Bay-Smidt, A. M., Olsen, C. E., & Møller, B. L. (1998). Analysis of starch-bound glucose 3-phosphate and glucose 6-phosphate using controlled acid treatment combined with high-performance anion-exchange chromatography. *Journal of Chromatography A, 829*(1–2), 385-391.

Blennow, A., & Engelsen, S. B. (2010). Helix-breaking news: fighting crystalline starch energy deposits in the cell. *Trends in Plant Science, 15*(4), 236-240.

Blennow, A., Nielsen, T. H., Baunsgaard, L., Mikkelsen, R., & Engelsen, S. B. (2002). Starch phosphorylation: a new front line in starch research. *Trends Plant Sci, 7*(10), 445-450.

Blennow, A., Wischmann, B., Houborg, K., Ahmt, T., Jørgensen, K., Engelsen, S. B., Bandsholm, O., & Poulsen, P. (2005). Structure function relationships of transgenic starches with engineered phosphate substitution and starch branching. *International Journal of Biological Macromolecules, 36*(3), 159-168.

Buléon, A., Colonna, P., Planchot, V., & Ball, S. (1998). Starch granules: structure and biosynthesis. *International Journal of Biological Macromolecules, 23*(2), 85-112.

Buleon, A., Cotte, M., Putaux, J. L., d'Hulst, C., & Susini, J. (2014). Tracking sulfur and phosphorus within single starch granules using synchrotron X-ray microfluorescence mapping. *Biochim Biophys Acta, 1840*(1), 113-119.

Carpenter, M., Joyce, N., Butler, R., Genet, R., & Timmerman-Vaughan, G. (2012). A mass spectrometric method for quantifying C3 and C6 phosphorylation of starch. *Analytical Biochemistry, 431*(2), 115-119.

Delvalle, D., Dumez, S., Wattebled, F., Roldan, I., Planchot, V., Berbezy, P., Colonna, P., Vyas, D., Chatterjee, M., Ball, S., Merida, A., & D'Hulst, C. (2005). Soluble starch synthase I: a major determinant for the synthesis of amylopectin in Arabidopsis thaliana leaves. *Plant J, 43*(3), 398-412.

Haebel, S., Hejazi, M., Frohberg, C., Heydenreich, M., & Ritte, G. (2008). Mass spectrometric quantification of the relative amounts of C6 and C3 position phosphorylated glucosyl residues in starch. *Analytical Biochemistry, 379*(1), 73-79.

Hansen, P. I., Spraul, M., Dvortsak, P., Larsen, F. H., Blennow, A., Motawia, M. S., & Engelsen, S. B. (2009). Starch phosphorylation—Maltosidic restrains upon 3′- and 6′-phosphorylation investigated by chemical synthesis, molecular dynamics and NMR spectroscopy. *Biopolymers, 91*(3), 179-193.

Hejazi, M., Fettke, J., Haebel, S., Edner, C., Paris, O., Frohberg, C., Steup, M., & Ritte, G. (2008). Glucan, water dikinase phosphorylates crystalline maltodextrins and thereby initiates solubilization. *The Plant Journal, 55*(2), 323-334.

Hejazi, M., Mahlow, S., & Fettke, J. (2014). The glucan phosphorylation mediated by α-glucan, water dikinase (GWD) is also essential in the light phase for a functional transitory starch turn-over. *Plant Signaling & Behavior, 9*(7), e28892.

Hizukuri, S., Tabata, S., Kagoshima, & Nikuni, Z. (1970). Studies on Starch Phosphate Part 1. Estimation of glucose-6-phosphate residues in starch and the presence of other bound phosphate(s). *Starch - Stärke, 22*(10), 338-343.

Ithaca, T. J. S., & Maywald, E. C. (1968). Some Unusual Properties of Pakistan 'Shoti' Starch. *Starch - Stärke, 20*(11), 362-365.

Jacobsen, H. B., Madsen, M. H., Christiansen, J., & Nielsen, T. H. (1998). The degree of starch phosphorylation as influenced by phosphate deprivation of potato (Solanum tuberosum L.) plants. *Potato Research, 41*(2), 109-116.

Kasemsuwan, T., & Jane, J. L. (1996). Quantitative method for the survey of starch phosphate derivatives and starch phospholipids by 31P nuclear magnetic resonance spectroscopy. *Cereal chemistry, 73*(6), 702-707.

Kötting, O., Pusch, K., Tiessen, A., Geigenberger, P., Steup, M., & Ritte, G. (2005). Identification of a Novel Enzyme Required for Starch Metabolism in Arabidopsis Leaves. The Phosphoglucan, Water Dikinase. *Plant Physiology, 137*(1), 242-252.

Kozlov, S. S., Blennow, A., Krivandin, A. V., & Yuryev, V. P. (2007). Structural and thermodynamic properties of starches extracted from GBSS and GWD suppressed potato lines. *International Journal of Biological Macromolecules, 40*(5), 449-460.

Mikkelsen, R., Baunsgaard, L., & Blennow, A. (2004). Functional characterization of alpha-glucan,water dikinase, the starch phosphorylating enzyme. *Biochem J, 377*(Pt 2), 525-532.

Morrison, W. R. (1964). A fast, simple and reliable method for the microdetermination of phosphorus in biological materials. *Analytical Biochemistry, 7*(2), 218-224.

Muhrbeck, P., & Tellier, C. (1991). Determination of the Posphorylation of Starch from Native Potato Varieties by 31P NMR. *Starch - Stärke, 43*(1), 25-27.

O'Shea, M. G., Samuel, M. S., Konik, C. M., & Morell, M. K. (1998). Fluorophore-assisted carbohydrate electrophoresis (FACE) of oligosaccharides: efficiency of labelling and high-resolution separation. *Carbohydrate Research, 307*(1–2), 1-12.

Ritte, G., Heydenreich, M., Mahlow, S., Haebel, S., Kötting, O., & Steup, M. (2006). Phosphorylation of C6- and C3-positions of glucosyl residues in starch is catalysed by distinct dikinases. *FEBS Letters, 580*(20), 4872-4876.

Santelia, D., Kotting, O., Seung, D., Schubert, M., Thalmann, M., Bischof, S., Meekins, D. A., Lutz, A., Patron, N., Gentry, M. S., Allain, F. H., & Zeeman, S. C. (2011). The phosphoglucan phosphatase like sex Four2 dephosphorylates starch at the C3-position in Arabidopsis. *Plant Cell, 23*(11), 4096-4111.

Skeffington, A. W., Graf, A., Duxbury, Z., Gruissem, W., & Smith, A. M. (2014). Glucan, Water Dikinase Exerts Little Control over Starch Degradation in Arabidopsis Leaves at Night. *Plant Physiology, 165*(2), 866-879.

Tabata, S., & Hizukuri, S. (1971). Studies on Starch Phosphate. Part 2. Isolation of Glucose 3-Phosphate and Maltose Phosphate by Acid Hydrolysis of Potato Starch. *Starch - Stärke, 23*(8), 267-272.

Viksø-Nielsen, A., Blennow, A., Jørgensen, K., Kristensen, K. H., Jensen, A., & Møller, B. L. (2001). Structural, Physicochemical, and Pasting Properties of Starches from Potato Plants with Repressed r1-Gene†. *Biomacromolecules, 2*(3), 836-843.

Yu, T.-S., Kofler, H., Häusler, R. E., Hille, D., Flügge, U.-I., Zeeman, S. C., Smith, A. M., Kossmann, J., Lloyd, J., Ritte, G., Steup, M., Lue, W.-L., Chen, J., & Weber, A. (2001). The Arabidopsis sex1 Mutant Is Defective in the R1 Protein, a General Regulator of Starch Degradation in Plants, and Not in the Chloroplast Hexose Transporter. *The Plant Cell Online, 13*(8), 1907-1918.